\documentstyle[prb,aps,eqsecnum]{revtex}
\begin{document}
\draft
\title {Properties of some mean-field like approximations for the
triangular Ising antiferromagnet}
\author {Alessandro Pelizzola and Marco Pretti}
\address {Istituto Nazionale di Fisica della Materia and Dipartimento 
di Fisica del Politecnico di Torino, I-10129 Torino, Italy}
\date{\today}
\maketitle
\begin{abstract}
Motivated by a recent proposal of a Bethe approximation for the
triangular Ising antiferromagnet [Phys. Rev. B {\bf 56}, 8241 (1997)],
which seems to predict a disordered phase at any temperature in zero
field, 
we analyze in some detail several mean-field like approximations for
this model, namely the Bethe approximation itself, the cluster variation
method and the hard-spin mean-field theory. We show: (i) that the
disordered phase predicted by the Bethe approximation is unphysical at
low enough temperature because of a negative entropy; (ii) how the
results of the cluster variation method (namely, zero temperature
entropy and critical temperature of the spurious transition) converge
to the exact ones for increasing cluster size; (iii) that it is
possible to construct a cluster variation approximation which yields a
disordered phase which is stable down to zero temperature; (iv) a few, so far
unknown, zero temperature results (entropy and internal energy) of the
hard-spin mean-field theory.
\end{abstract}

\section{Introduction}
The antiferromagnetic Ising model on the triangular lattice 
can be considered as the prototype of frustrated lattice models. 
It is defined by the (reduced) hamiltonian
\begin{equation} 
  {\cal H}/|J| = \sum_{\langle i j \rangle} \sigma_i \sigma_j
  - h \sum_i \sigma_i,
\end{equation} 
where $J<0$ is the antiferromagnetic interaction,
$h = H/|J|$ is the reduced uniform magnetic field and
$\sigma_i$ is an Ising spin at site $i$,
which can take the values $+1$ and $-1$.
The former sum runs over nearest neighbor (NN) pairs
and the latter runs over all lattice sites.
In zero field the ground state of the model is highly frustrated
since the minimization of bond energies would require
that all pairs of NN spins are antiparallel,
but this condition cannot be fulfilled even on a simple triangle.
On the contrary,
six out of the eight possible configurations of a triangular plaquette
(that is, all except the two fully parallel ones)
are selected as the lowest energy ones,
and this yields a huge degeneracy of the global ground state.
In zero field, the model has been solved exactly~\cite{wannier,houtappel}
and the solution shows that 
the disordered phase is stable down to zero temperature and 
the zero temperature entropy per site (in the thermodynamic limit) is
$S/N \approx 0.323066 \times k_B$,
where $N$ is the number of sites
and $k_B$ is the Boltzmann constant. 

Although an exact solution is available, a lot of work has focused on
mean-field approximations, since the present model can be viewed as a
playground for testing approximate methods against exact results
before applying them to more general and difficult models. The usual
(single site) mean-field approximation fails qualitatively to predict
the zero field behavior of the model: the disordered phase is
thermodynamically stable only down to a certain transition
temperature, below which one finds a three-sublattice ordered
phase. Also the more sophisticated Cluster Variation Method (CVM)
\cite{ktk,kkk} gives the same qualitative picture, although with a lower
transition temperature, which decreases upon increasing the cluster
size. An interesting approach, 
which seems to display a qualitatively correct behavior,
is the Hard-Spin Mean-Field (HSMF) theory by Berker and coworkers
\cite{berker1,berker2,maritan,berker3,berker4}, although quantities like
entropy and free energy are to be calculated according to recipes which
are not very well defined~\cite{berker4}. 
Furthermore, a new investigation based on
the Bethe-Peierls (BP) approximation has been recently carried out by
Tamashiro and Salinas~\cite{salinas}: the authors claim that this
approximation predicts a paramagnetic phase which in zero
magnetic field is stable down to zero temperature. 

The purpose of the present paper is twofold: we shall first show that
the paramagnetic phase found by Tamashiro and Salinas~\cite{salinas} is
unphysical at low temperatures, by explicitly calculating the entropy,
which turns out to be negative; then we shall address the more general
issue of analyzing carefully the other approximations, with a
particular attention to the zero field, zero temperature case. In the
case of the CVM we shall show that the zero field, zero temperature
entropy of the disordered phase converges, for increasing cluster
size, to the exact value according to a power law, while the critical
temperature converges very slowly to the exact value of zero;
furthermore we shall shortly discuss a CVM 
approximation which does not predict a spurious transition in zero
field. In the case of the HSMF theory we shall give a well-defined
recipe to calculate the free energy and the entropy and use it to
evaluate the zero field, zero temperature entropy.

\section{The Bethe-Peierls approximation}

The BP approximation~\cite{burley} has long been used
to improve upon the ordinary mean-field results.
It is the simplest mean-field like approximation
which takes into account two-site correlations.

Recently, Tamashiro and Salinas~\cite{salinas} applied the BP
approximation to the present model, splitting the triangular lattice
in the natural way into its three sublattices. They report that in the
case of vanishing magnetic field the disordered (paramagnetic) phase
is stable down to zero temperature. This is actually what the
approximation predicts, but a deeper analysis is in order to
understand what kind of disordered phase we are talking of at low
temperature.
Since in the next section
we are going to discuss the CVM, it is worth mentioning that the CVM
itself reduces, if one chooses a NN pair as the maximal
cluster, to the BP approximation.
Let us consider the reduced entropy density $s=S/Nk_B$ given by the
CVM for the triangular lattice 
with the assumption of three non equivalent sublattices (say $A,B,C$)
and the above choice of basic clusters (see e.g. \cite{An}):
\begin{eqnarray}
s & = & - \left[ {\rm Tr}(\rho_{AB} \ln \rho_{AB}) + 
                 {\rm Tr}(\rho_{BC} \ln \rho_{BC}) + 
                 {\rm Tr}(\rho_{CA} \ln \rho_{CA}) \right] 
\nonumber \\ && + \frac{5}{3} 
          \left[ {\rm Tr}(\rho_{A} \ln \rho_{A}) + 
                 {\rm Tr}(\rho_{B} \ln \rho_{B}) + 
                 {\rm Tr}(\rho_{C} \ln \rho_{C}) \right].
\end{eqnarray}
Here $\rho_{\alpha \alpha^\prime}$ is the NN pair density matrix
(which is diagonal since we are dealing with a classical model)
and its diagonal elements
$\rho_{\alpha \alpha^\prime}(\sigma,\sigma^\prime)$ represent 
the probability of finding the NN spins belonging to the sublattices 
$\alpha$ and $\alpha^\prime$ in the states $\sigma$ and $\sigma^\prime$, 
respectively.
Similarly $\rho_\alpha = {\rm Tr}_{\alpha'}(\rho_{\alpha \alpha^\prime})$
is the reduced (single site) density matrix,
whose diagonal elements $\rho_\alpha (\sigma)$ represent 
the probability of finding a spin belonging to the sublattice 
$\alpha$ in the state $\sigma$.
Rewriting these probabilities in terms of the magnetizations
$m_\alpha = \langle \sigma_\alpha \rangle$
and the NN pair correlations
$c_{\alpha\alpha'} = \langle \sigma_\alpha \sigma_{\alpha'} \rangle$
(where $\sigma_\alpha$ and $\sigma_\alpha'$ denote NN spin variables
on the sublattices $\alpha$ and $\alpha'$, respectively
and as usual $\langle\cdot\rangle$ denotes thermal average) one obtains
\begin{equation}
  s = -\sum_{\langle \alpha \alpha' \rangle}
  \sum_{\sigma = \pm 1} \sum_{\sigma' = \pm 1} {\cal L}
  \left(
    \frac{1 + \sigma m_\alpha
            + \sigma' m_{\alpha'}
            + \sigma\sigma' c_{\alpha\alpha'} }{4}
  \right)
  +\frac{5}{3} \sum_{\alpha} \sum_{\sigma = \pm 1} {\cal L}
    \left(
      \frac{1 + \sigma m_\alpha}{2}
    \right),
  \label{entropia_BP}
\end{equation}
where the outer sum runs over sublattice pairs
($\alpha\alpha' = AB, BC, CA$) in the former term,
over sublattices ($\alpha = A, B, C$) in the latter
and we have defined ${\cal L}(x) = x \ln x$.
The (reduced) internal energy density 
(which more rigorously we shall refer to as an enthalpy)
$\epsilon = \langle {\cal H} \rangle/N|J|$
can be written exactly as
\begin{equation}
  \epsilon = \sum_{\langle \alpha \alpha' \rangle} c_{\alpha\alpha'} 
  - \frac{1}{3} \sum_{\alpha} h m_\alpha.
  \label{energia_interna_BP}
\end{equation}
The minimization of the (reduced) free energy density
\begin{equation}
  g = G/N|J| = \epsilon - ts
  \label{energia_libera}
\end{equation}
(where $t=k_BT/|J|$ and $T$ is the absolute temperature)
with respect to the magnetizations and the NN correlations 
can then be performed. 
We introduce the conditioned average of a spin (on sublattice $\alpha$)
with respect to its NN (on sublattice $\alpha'$)
\begin{equation}
  m_{\alpha|\alpha'}(\sigma) =
  \left. \left\langle \sigma_\alpha \right\rangle \frac{}{} \right|
  _{\displaystyle \sigma_{\alpha'} = \sigma} =
  \sum_{\sigma = \pm 1} \sigma 
  \frac{\rho_{\alpha\alpha'}(\sigma,\sigma')}{\rho_{\alpha'}(\sigma')},
\end{equation}
which is related to the previous variables by the identity:
\begin{equation}
  m_\alpha + \sigma c_{\alpha\alpha'} =
  m_{\alpha|\alpha'}(\sigma)(1 + \sigma m_{\alpha'})
  \hspace{1cm} \sigma = \pm 1.
\end{equation}
The stationarity conditions for $g$
can thus be written as
\begin{eqnarray}
  \frac{\partial g}{\partial m_\alpha} & = &
  -\frac{1}{3}h +\frac{1}{3}t\tanh^{-1}m_\alpha
  +\frac{1}{4}t\sum_{\alpha'\neq\alpha}
  \ln\frac{1-m_{\alpha'|\alpha}^2(+)}{1-m_{\alpha'|\alpha}^2(-)} = 0
  \label{derivate_m} \\
  \frac{\partial g}{\partial c_{\alpha\alpha'}} & = &
  1 + t\,\frac{\tanh^{-1}m_{\alpha|\alpha'}(+) 
              -\tanh^{-1}m_{\alpha|\alpha'}(-)}{2} = 0,
  \label{derivate_c}
\end{eqnarray}
where the sum involves two sublattices only
(for instance, if $\alpha=A$ then $\alpha'=B,C$).
If the following quantity is defined
\begin{equation}
  \beta\eta_{\alpha\alpha'} \stackrel{\triangle}{=}
  \frac{\tanh^{-1}m_{\alpha|\alpha'}(+) 
       +\tanh^{-1}m_{\alpha|\alpha'}(-)}{2}
\end{equation}
(being as usual $\beta=1/k_BT$),
the equations \ref{derivate_c} become
\begin{equation}
  m_{\alpha|\alpha'}(\sigma) =
  \tanh\left(\beta\eta_{\alpha\alpha'} - \sigma\beta|J|\right)
  \hspace{1cm} \sigma = \pm 1.
  \label{mediecond-campieff}
\end{equation}
Substituting into Eq.~\ref{derivate_m} 
one obtains, with a bit of algebra:
\begin{equation}
  \beta|J|\frac{\partial g}{\partial m_\alpha} =
  -\frac{1}{3}\beta H +\frac{1}{3}\tanh^{-1}m_\alpha
  +\sum_{\alpha'\neq\alpha} \tanh^{-1}\left[
    \tanh\left(\beta|J|\right)\tanh\left(\beta\eta_{\alpha'\alpha}\right)
  \right] = 0.
\end{equation}
This last equation, 
together with Eqs.~\ref{mediecond-campieff},
is equivalent to the self consistent equation
derived in Ref.~\onlinecite{salinas},
and $\eta_{\alpha\alpha'}$ turn out to be
the same effective fields as those of the BP approximation.
Once the equivalence is proved we can go on working
in the CVM approach.
In the uniform case
($m_\alpha = m \; \forall \alpha$ and
$c_{\alpha\alpha'} = c \; \forall \alpha\neq\alpha'$)
the stationarity conditions for $g$
can be easily written (without the above manipulations)
in the following form:
\begin{eqnarray}
  m & = & \pm \frac{1}{2} \sqrt{(1+c)^2-(1-c)^2e^{-4/t}}
  \label{m-tc} \\
  h & = & t \left[3\tanh^{-1}\frac{2m}{1+c}-5\tanh^{-1}m\right].
  \label{h-tmc}
\end{eqnarray}
Eq.~\ref{m-tc} allows us to evaluate the magnetization $m$
as a function of the pair correlation $c$ and the reduced temperature $t$.
This equation, in the limit $t \to 0$,
states that a disordered phase ($m=0$) can only have $c=-1$.
The zero temperature entropy and enthalpy 
(the latter coinciding with the internal energy $u=U/N|J|$ if $h=0$)
can then be evaluated by
Eqs.~\ref{entropia_BP} and \ref{energia_interna_BP}
and turn out to be respectively $s=-2\ln 2$ 
and (being $h=0$) $\epsilon=u=3c=-3$.
The same results can be obtained by the BP free energy
derived in Ref.~\onlinecite{salinas}.
If $h=0$ then the magnetization (and hence the local effective
field) of the paramagnetic phase vanish, which corresponds to taking
$x_1 = x_2 = 0$ in Eq. 24 of Ref.~\onlinecite{salinas}.
In this case the free energy reduces to (using $v = \tanh(\beta J)$)
\begin{equation}
  \beta|J|g = - \ln 2 + \frac{3}{2} \ln (1 - v^2)
  \stackrel{\beta\to\infty}{\sim}  2 \ln 2 - 3 \beta |J|,
\end{equation}
where the $\beta\to\infty$ asymptotic expression clearly agrees
with our previous discussion.
A couple of comments on these results are in order.
A pair correlation equal to $-1$ means that {\em every} pair of
NN spins is, in the ground state, in an antiparallel
state. This is of course not possible, as mentioned in the
Introduction. The BP approximation, taking into account only NN pairs
and single sites, completely neglects frustration effects and predicts
an internal energy (and hence, at low enough temperatures, a free
energy) which is much lower than the exact one, $u_{\rm ex} = -1$
\cite{wannier,houtappel} 
thus ``stabilizing'' the disordered phase. We note in 
passing that the BP approximation usually gives 
a free energy which is an upper bound to the exact one. 

It is clear that the state which is predicted to be the ground state
cannot exist, and this is reflected in the {\em negative} value of the
zero temperature entropy. Of course, such unphysical effects
will not be limited to the point $(t=0,h=0)$, but will be found in
a finite region of the phase diagram around this point. In order to
give an idea of the size of this region we only mention that the zero
field entropy vanishes at a temperature $t \approx 1.28$,
and is negative below this temperature. Another, more
stringent, criterion, can be based on the NN correlation $c$:
a NN pair probability distribution can be obtained
by partial trace from a triangle probability distribution only if 
$c \ge -1/3$. In the
BP approximation $c = -1$ in the ground state and $c = -1/3$ at a
temperature $t = 2/\ln 2 \approx 2.89$. Below such a temperature the BP
results are certainly unphysical, since they predict a NN pair probability
distribution which cannot exist in a triangular lattice.
In Fig. \ref{fig:ccost}
we report the curves defined by $s=0$ and $c=-1/3$ in the $(h,t)$
plane.
The former is evaluated numerically by Eq.~\ref{entropia_BP}
while the latter can be analytically drawn
by substituting Eq.~\ref{m-tc} into Eq.~\ref{h-tmc}.

\section{The Cluster Variation Method}

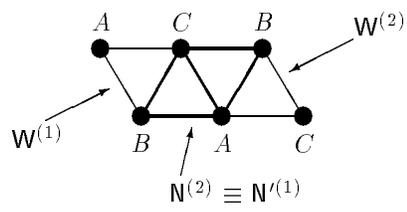
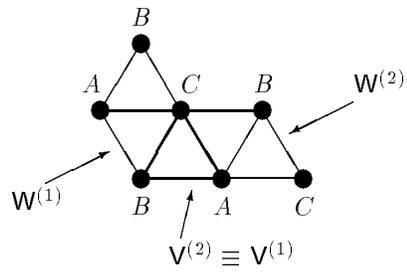
\begin{figure}
\caption{
  Examples of compatibility conditions.
  (a) For $L=2$ it is shown that the cluster obtained by
  the superposition of ${\sf W}^{(2)}$ ($BCABC$)
  and ${\sf W}^{(1)}$ ($ABCAB$)
  is ${\sf N}^{(2)}$ ($BCAB$, thick lines);
  hence the same probability distribution $\rho_{\sf N}^{(2)}$
  must be obtained
  either by tracing $\rho_{\sf W}^{(2)}$ over site $5$ ($C$)
  or $\rho_{\sf W}^{(1)}$ over site $1$ ($A$).
  (b) Again for $L=2$ it is shown that the cluster obtained by
  the superposition of ${\sf W}^{(2)}$ ($BCABC$)
  and a rotated ${\sf W}^{(1)}$ ($ABCAB$)
  is ${\sf V}^{(2)}$ ($BCA$, thick lines),
  coinciding with ${\sf V}^{(1)}$ ($ABC$);
  hence the same probability distribution $\rho_{\sf V}^{(2)}$
  must be obtained
  either by tracing $\rho_{\sf W}^{(2)}$ over sites $4,5$ ($B,C$)
  or $\rho_{\sf W}^{(1)}$ over sites $4,5$ ($A,B$).
  It is easy to generalize both examples to any cluster size $L$.
  The same constraints must be satisfied by any pair
  (${\sf W}^{(2)}{\sf W}^{(1)}$,
   ${\sf W}^{(3)}{\sf W}^{(2)}$,
   ${\sf W}^{(1)}{\sf W}^{(3)}$):
  conditions (a) imply a translational invariance
  and conditions (b) a rotational invariance.
  }
  \label{fig:vincoli}
\end{figure}

In this section we develop a CVM approximation
with the usual assumption of three non equivalent sublattices.
We increase the cluster size anisotropically (just along one direction),
according to Kikuchi's ${\rm B}_{2L}$-hierarchy,
which has been shown to converge
to the exact solution~\cite{Kikuchi-Brush}.
The proof of convergence has been made rigorous
in the special case of
the Ising model on a square lattice~\cite{Schlijper}.
This approach is more convenient,
from the point of view of computational efficiency,
than choosing larger basic clusters in both directions,
as in Refs.~\onlinecite{ktk,kkk}.
In our approximation
the basic clusters are composed by $L$ (downward pointing) triangles
and contain $2L+1$ lattice points,
labelled as in Fig.~\ref{fig:indici}.
Due to their shape we will denote such clusters as {\sf W}-class.
As usual the hamiltonian can be written as a sum of cluster hamiltonians:
\begin{equation}
  {\cal H}/|J| = \sum_W
  h_{\sf W} \left( \mbox{\boldmath $\sigma$}_{1,2,\dots,2L+1}^{(W)} \right),
\end{equation}
where the sum runs over all basic clusters
and $\mbox{\boldmath $\sigma$}_{1,2,\dots,2L+1}^{(W)}$
is the spin configuration of the current cluster. 
>From now on we shall use the following notation:
\begin{equation}
  \mbox{\boldmath $\sigma$}_{i_1,i_2,\dots,i_n}
  \stackrel{\triangle}{=}
  \left\{ \sigma_{i_1}, \sigma_{i_2}, \dots, \sigma_{i_n} \right\},
\end{equation}
indicating a set of spins.
The cluster hamiltonian has the following expression:
\begin{equation}
  h_{\sf W} \left( \mbox{\boldmath $\sigma$}_{1,2,\dots,2L+1} \right) 
  = \frac{1}{  2L   } \sum_{i=1}^{2L  } \sigma_{i} \sigma_{i+1}
  + \frac{1}{2(2L-1)} \sum_{i=1}^{2L-1} \sigma_{i} \sigma_{i+2} 
  - \frac{h}{2(2L+1)} \sum_{i=1}^{2L+1} \sigma_{i},
\end{equation}
where $\sigma_i$ is the $i$-th spin (see Fig.~\ref{fig:indici}),
while the first sum runs over oblique bonds,
the second over horizontal bonds
and the third over sites.
The coefficients before each sum avoid multiple countings:
every oblique bond is shared by $2L$ basic clusters,
every horizontal bond by $2(2L-1)$ 
and every site by $2(2L+1)$ basic clusters.
In the hypothesis of three non equivalent sublattices
(labelled $A$, $B$ and $C$)
three non equivalent basic clusters
(${\sf W}^{(1)}$, ${\sf W}^{(2)}$ and ${\sf W}^{(3)}$,
where superscripts denote different spin probability distributions)
must be taken into account,
as shown in Fig.~\ref{fig:cluster} (first row).
Observing that the same fraction ($1/3$) of each cluster type is present,
the enthalpy can be written as
\begin{equation}
  \epsilon = \frac{2}{3} \sum_{\kappa=1}^{3} \; \;
  \sum_{\mbox{\boldmath $\sigma$}_{1,2,\dots,2L+1}}
  \rho_{\sf W}^{(\kappa)}
  \left(\mbox{\boldmath $\sigma$}_{1,2,\dots,2L+1}\right)
  h_{\sf W}\left(\mbox{\boldmath $\sigma$}_{1,2,\dots,2L+1}\right),
  \label{energia_interna_CVM}
\end{equation}
where the $\kappa$ index scans cluster types and
$\rho_{\sf W}^{(\kappa)}
\left(\mbox{\boldmath $\sigma$}_{1,2,\dots,2L+1}\right)$
is the probability of the spin configuration
$\mbox{\boldmath $\sigma$}_{1,2,\dots,2L+1}$
of the cluster ${\sf W}^{(\kappa)}$
(the inner sum runs over all possible configurations).
Notice that the coefficient $2$ is the total number of ${\sf W}$-clusters
per lattice site.
Following the most recent formulation of the CVM~\cite{An}
the entropy can be easily written
as a linear combination (with suitable coefficients) of cluster entropies
relative to a set of basic clusters
(${\sf W}^{(1)}$, ${\sf W}^{(2)}$, ${\sf W}^{(3)}$ in our case)
and their subclusters.
For a given cluster $\gamma$ the coefficient is obtained by
the product of a number $a_\gamma$,
evaluated by Moebius inversion~\cite{An},
and the ratio $N_\gamma / N$
(number of $\gamma$-clusters per lattice site).
It turns out that in our problem
only three subcluster classes
have non vanishing $a_\gamma$.
They are displayed in Fig.~\ref{fig:cluster}
(labelled by ${\sf N}$, ${\sf O}$, ${\sf E}$)
while the corresponding coefficients
are given in Tab.~\ref{tab:coefficienti}.
We finally obtain
\begin{eqnarray}
  s & = & - \frac{1}{3} \sum_{\kappa=1}^{3} \; \;
  \sum_{\mbox{\boldmath $\sigma$}_{1,2,\dots,2L+1}}
  \rho_{\sf W}^{(\kappa)}
  \left(\mbox{\boldmath $\sigma$}_{1,2,\dots,2L+1}\right)
  \left[
   2\ln\rho_{\sf W}^{(\kappa)}
   \left(\mbox{\boldmath $\sigma$}_{1,2,\dots,2L+1}\right)
   \right. \nonumber \\ & & \left.
  - 2\ln\rho_{\sf N}^{(\kappa)}
    \left(\mbox{\boldmath $\sigma$}_{1,2,\dots,2L  }\right)
  - \ln\rho_{\sf O}^{(\kappa)}
    \left(\mbox{\boldmath $\sigma$}_{1,3,\dots,2L+1}\right)
  + \ln\rho_{\sf E}^{(\kappa)}
    \left(\mbox{\boldmath $\sigma$}_{2,4,\dots,2L  }\right)
  \right],
  \label{entropia_CVM}
\end{eqnarray}
where the subcluster probability distributions have been defined:
\begin{eqnarray}
  \rho_{\sf N}^{(\kappa)}
  \left(\mbox{\boldmath $\sigma$}_{1,2,\dots,2L  }\right)
  & = & \sum_{\sigma_{2L+1}}
  \rho_{\sf W}^{(\kappa)}
  \left(\mbox{\boldmath $\sigma$}_{1,2,\dots,2L+1}\right)
  \nonumber \\
  \rho_{\sf O}^{(\kappa)}
  \left(\mbox{\boldmath $\sigma$}_{1,3,\dots,2L+1}\right)
  & = & \sum_{\mbox{\boldmath $\sigma$}_{2,4,\dots,2L  }}
  \rho_{\sf W}^{(\kappa)}
  \left(\mbox{\boldmath $\sigma$}_{1,2,\dots,2L+1}\right)
  \nonumber \\
  \rho_{\sf E}^{(\kappa)}
  \left(\mbox{\boldmath $\sigma$}_{2,4,\dots,2L  }\right)
  & = & \sum_{\mbox{\boldmath $\sigma$}_{1,3,\dots,2L+1}}
  \rho_{\sf W}^{(\kappa)}
  \left(\mbox{\boldmath $\sigma$}_{1,2,\dots,2L+1}\right).
\end{eqnarray}
It is then evident that the free energy $g = \epsilon - ts$
is a function of the probability distributions
of the basic clusters only.
The variational procedure, with respect to these variables,
has yet to satisfy three kinds of constraints:
\begin{enumerate}
\item
the normalization conditions ($3$ constraints)
\begin{equation}
  \sum_{\mbox{\boldmath $\sigma$}_{1,2,\dots,2L+1}}
  \rho_{\sf W}^{(\kappa)}
  \left(\mbox{\boldmath $\sigma$}_{1,2,\dots,2L+1}\right)
  = 1 \hspace{1cm} \kappa=1,2,3;
  \label{normalizzazione}
\end{equation}
\item
the ``translational'' compatibility conditions
($3 \times 2^{2L}$ constraints)
\begin{equation}
  \rho_{\sf N }^{(\kappa  )}
  \left(\mbox{\boldmath $\sigma$}_{1,2,\dots,2L}\right) =
  \rho_{\sf N'}^{(\kappa-1)}
  \left(\mbox{\boldmath $\sigma$}_{1,2,\dots,2L}\right)
  \hspace{1cm}
  \forall \mbox{\boldmath $\sigma$}_{1,2,\dots,2L}
  \; ; \; \; \kappa=1,2,3
  \label{comp_trasl}
\end{equation}
where
\begin{equation}
  \rho_{\sf N'}^{(\kappa)}
  \left(\mbox{\boldmath $\sigma$}_{2,3,\dots,2L+1}\right)
  = \sum_{\sigma_1}
  \rho_{\sf W }^{(\kappa)}
  \left(\mbox{\boldmath $\sigma$}_{1,2,\dots,2L+1}\right)
\end{equation}
and the $\kappa$ indices are understood modulo $3$
(as in following occurrences);
\item
the ``rotational'' compatibility conditions
($3 \times 2^3$ constraints)
\begin{equation}
  \rho_{\sf V}^{(\kappa  )}\left(\mbox{\boldmath $\sigma$}_{1,2,3}\right) =
  \rho_{\sf V}^{(\kappa-1)}\left(\mbox{\boldmath $\sigma$}_{3,1,2}\right)
  \hspace{1cm}
  \forall \mbox{\boldmath $\sigma$}_{1,2,3}
  \; ; \; \; \kappa=1,2,3
  \label{comp_rot}
\end{equation}
where
\begin{equation}
  \rho_{\sf V}^{(\kappa)}\left(\mbox{\boldmath $\sigma$}_{1,2,3}\right)
  = \sum_{\mbox{\boldmath $\sigma$}_{4,5,\dots,2L+1}}
  \rho_{\sf W}^{(\kappa)}
  \left(\mbox{\boldmath $\sigma$}_{1,2,\dots,2L+1}\right).
\end{equation}
\end{enumerate}
The last two constraints can be more easily understood
by the examples shown in Fig.~\ref{fig:vincoli}.
Notice that a new subcluster class has been defined,
namely ${\sf V}$ (the class of simple triangular plaquettes),
while, due to constraints,
${\sf N'}$ is actually the same class as ${\sf N}$.
In order to take into account the above described constraints
we have to introduce the Lagrange multipliers
and define the free energy functional
\begin{eqnarray}
  \tilde{g} & = g &
  - \frac{1}{3} \sum_{\kappa=1}^{3} 
    \mu^{(\kappa)}
    \sum_{\mbox{\boldmath $\sigma$}_{1,2,\dots,2L+1}}
    \rho_{\sf W}^{(\kappa)}
    \left(\mbox{\boldmath $\sigma$}_{1,2,\dots,2L+1}\right)
  \nonumber \\
  & & 
  - \frac{1}{3} \sum_{\kappa=1}^{3} \; \;
    \sum_{\mbox{\boldmath $\sigma$}_{1,2,\dots,2L  }}
    \upsilon^{(\kappa)}\left(\mbox{\boldmath $\sigma$}_{1,2,\dots,2L}\right)
    \left[
      \rho_{\sf N }^{(\kappa  )}
      \left(\mbox{\boldmath $\sigma$}_{1,2,\dots,2L}\right) -
      \rho_{\sf N'}^{(\kappa-1)}
      \left(\mbox{\boldmath $\sigma$}_{1,2,\dots,2L}\right)
    \right]
  \nonumber \\
  & & 
  - \frac{1}{3} \sum_{\kappa=1}^{3} \; \;
    \sum_{\mbox{\boldmath $\sigma$}_{1,2,3}}
    \omega^{(\kappa)} \left(\mbox{\boldmath $\sigma$}_{1,2,3}\right)
    \left[
      \rho_{\sf V}^{(\kappa  )}\left(\mbox{\boldmath $\sigma$}_{1,2,3}\right) -
      \rho_{\sf V}^{(\kappa-1)}\left(\mbox{\boldmath $\sigma$}_{3,1,2}\right)
    \right],
\end{eqnarray}
where $g$ is defined by Eqs.~\ref{energia_libera},
\ref{energia_interna_CVM} and \ref{entropia_CVM},
while $\mu^{(\kappa)}$,
$\upsilon^{(\kappa)} \left(\mbox{\boldmath $\sigma$}_{1,2,\dots,2L}\right)$
and $\omega^{(\kappa)} \left(\mbox{\boldmath $\sigma$}_{1,2,3}\right)$
are the Lagrange multipliers related to constraints~\ref{normalizzazione}, 
\ref{comp_trasl}, \ref{comp_rot} respectively.
Taking the derivatives of $\tilde{g}$
with respect to the probabilities
$\rho_{\sf W}^{(\kappa)}
\left(\mbox{\boldmath $\sigma$}_{1,2,\dots,2L+1}\right)$
and setting them to zero,
one easily obtains the Natural Iteration (NI) equations~\cite{Kikuchi74}
\begin{eqnarray}
  \hat{\rho}_{\sf W}^{(\kappa)}
  \left(\mbox{\boldmath $\sigma$}_{1,2,\dots,2L+1}\right)
  & = &
  \exp \left[
    \mu^{(\kappa)} / 2
    -h_{\sf W}\left(\mbox{\boldmath $\sigma$}_{1,2,\dots,2L+1}\right)
  \right] \nonumber \\ & \times &
  \exp \left[  
    \upsilon^{(\kappa  )}
    \left(\mbox{\boldmath $\sigma$}_{1,2,\dots,2L  }\right)
   -\upsilon^{(\kappa+1)}
    \left(\mbox{\boldmath $\sigma$}_{2,3,\dots,2L+1}\right)
  \right] \nonumber \\ & \times &
  \exp \left[  
    \omega^{(\kappa  )}\left(\mbox{\boldmath $\sigma$}_{1,2,3}\right)
   -\omega^{(\kappa+1)}\left(\mbox{\boldmath $\sigma$}_{2,3,1}\right)
  \right] \nonumber \\ & \times &
    \rho_{\sf N}^{(\kappa)}
    \left(\mbox{\boldmath $\sigma$}_{1,2,\dots,2L  }\right)
    \left[ \frac
  { \rho_{\sf O}^{(\kappa)}
  \left(\mbox{\boldmath $\sigma$}_{1,3,\dots,2L+1}\right) }
  { \rho_{\sf E}^{(\kappa)}
  \left(\mbox{\boldmath $\sigma$}_{2,4,\dots,2L  }\right) }
    \right]^{1 / 2},
  \label{equazioni}
\end{eqnarray}
where the ``hat'' on the left hand side
denotes the evaluation of the probabilities
at the ``next step'' of the NI method.
The convergence is reached when
a suitable distance between the current step and the next step
turns out to be less than some tolerance.
In order to compare results from different $L$,
and hence with a different number of variables,
we have used the $\infty$-norm instead of the $1$-norm.
The stopping test is thus
\begin{equation}
  \max_{\kappa=1,2,3} \;
  \max_{\mbox{\boldmath $\sigma$}_{1,2,\dots,2L+1}}
  \left| \hat{\rho}_{\sf W}^{(\kappa)}
  \left(\mbox{\boldmath $\sigma$}_{1,2,\dots,2L+1}\right) -
  \rho_{\sf W}^{(\kappa)}
  \left(\mbox{\boldmath $\sigma$}_{1,2,\dots,2L+1}\right) \right|
  < \varepsilon,
  \label{test_arresto}
\end{equation}
where usually $\varepsilon = 10^{-9}$ has been chosen.
The entire set of Lagrange multipliers
can be determined at every step
by means of a nested iterative procedure~\cite{Kikuchi76}.
The general scheme of the method is as follows.
\begin{enumerate}
\item
Choose a set of guess values for
$\rho_{\sf W}^{(\kappa)}
\left(\mbox{\boldmath $\sigma$}_{1,2,\dots,2L+1}\right)$
and set to zero the Lagrange multipliers
$\upsilon^{(\kappa)}\left(\mbox{\boldmath $\sigma$}_{1,2,\dots,2L}\right)$
and $\omega^{(\kappa)}\left(\mbox{\boldmath $\sigma$}_{1,2,3}\right)$.
\item
Evaluate the unnormalized
$\left.\hat{\rho}_{\sf W}^{(\kappa)}
\left(\mbox{\boldmath $\sigma$}_{1,2,\dots,2L+1}\right)
\right| _{\displaystyle \mu^{(\kappa)} = 0}$
by means of Eqs.~\ref{equazioni}.
\item
Compute a new estimate of the multipliers $\upsilon$ and $\omega$
by means of the following equations~\cite{Kikuchi76}:
\begin{eqnarray}
  \hat{\upsilon}^{(\kappa  )}
  \left(\mbox{\boldmath $\sigma$}_{1,2,\dots,2L}\right)
  & = & \upsilon^{(\kappa  )}
  \left(\mbox{\boldmath $\sigma$}_{1,2,\dots,2L}\right)
  - b_\upsilon \ln \frac
  { \rho_{\sf N }^{(\kappa  )}
    \left(\mbox{\boldmath $\sigma$}_{1,2,\dots,2L}\right) }
  { \rho_{\sf N'}^{(\kappa-1)}
    \left(\mbox{\boldmath $\sigma$}_{1,2,\dots,2L}\right) }
  \nonumber \\
  \hat{\omega}^{(\kappa  )}\left(\mbox{\boldmath $\sigma$}_{1,2,3}\right)
  & = & \omega^{(\kappa  )}\left(\mbox{\boldmath $\sigma$}_{1,2,3}\right)
  - b_\omega \ln \frac
  { \rho_{\sf V}^{(\kappa  )}\left(\mbox{\boldmath $\sigma$}_{1,2,3}\right) }
  { \rho_{\sf V}^{(\kappa-1)}\left(\mbox{\boldmath $\sigma$}_{3,1,2}\right) },
\end{eqnarray}
where $b_\upsilon$ and $b_\omega$ must be empirically chosen
in order to ``stabilize'' the numerical procedure
(we have found that $0.3$ is a good value for both parameters).
\item
Repeat steps 2 and 3 until convergence is reached
for the Lagrange multipliers
(with a stopping test similar to (\ref{test_arresto})
but generally a different tolerance $\varepsilon'$).
\item
Evaluate $\mu^{(\kappa)}$ according to the normalization conditions
(Eqs.~\ref{normalizzazione}),
which, applied to Eqs.~\ref{equazioni}, give
\begin{equation}
  \sum_{\mbox{\boldmath $\sigma$}_{1,2,\dots,2L+1}}
  \left.
  \hat{\rho}_{\sf W}^{(\kappa)}
  \left(\mbox{\boldmath $\sigma$}_{1,2,\dots,2L+1}\right)
  \right| _{\mu^{(\kappa)} = 0}
  = \exp \left(- \mu^{(\kappa)} / 2 \right)
\end{equation}
and determine correctly normalized probabilities.
\item
Repeat steps 2-5 until convergence is reached for the probabilities,
according to the stopping test (\ref{test_arresto}).
\end{enumerate}
Finally, when convergence is reached,
the equilibrium free energy can be easily evaluated as
\begin{equation}
  g = \frac{1}{3} \sum_{\kappa=1}^{3} \mu^{(\kappa)}.
\end{equation}

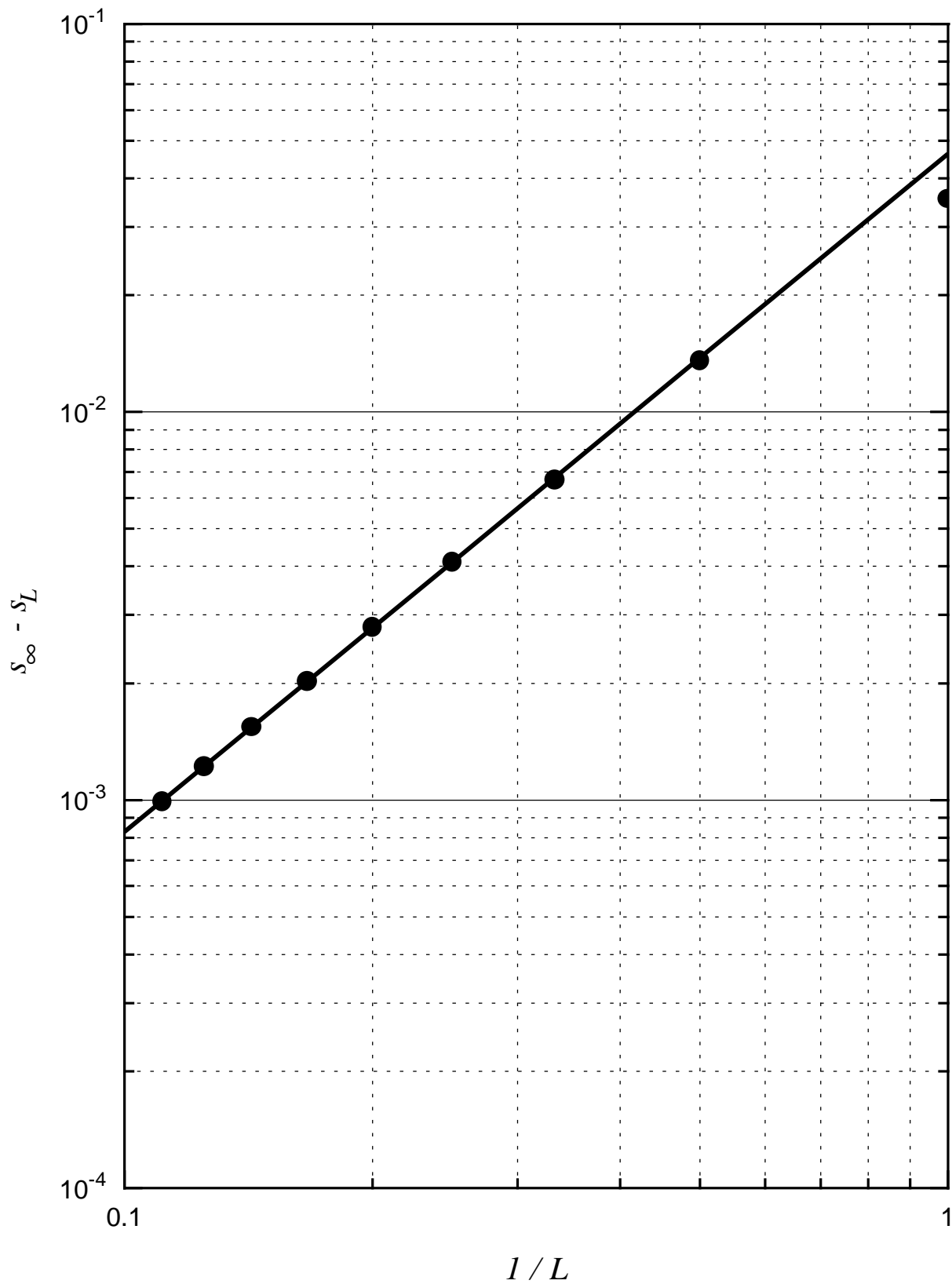
\begin{figure}
\caption{
  The difference between 
  the extrapolated zero temperature zero field entropy 
  $s_\infty \approx s_{\rm ex}$
  and that calculated in the CVM approximation $s_L$
  (for the disordered phase)
  is plotted versus the reciprocal cluster size $1/L$.
  Dots represent data obtained by the CVM
  while the solid straight line is 
  the least square fitting (data point $L=1 \div 3$ discarded).
  The slope of the straight line is the exponent $\psi$.
  }
\label{fig:entropia}
\end{figure}

We have carried out the investigation
increasing the cluster size parameter $L$,
as far as convergence could be reached 
with a reasonable computational effort
($L=9$ for the disordered phase and $L=8$ for the ordered phases).
It will be clear in the following that
it is convenient to rewrite the sublattice magnetizations
$m_A,m_B,m_C$ as follows:
\begin{eqnarray}
  m_A & = & {\rm Re}\left\{\overline{m}+Me^{i\varphi}\right\} 
  \nonumber \\
  m_B & = & {\rm Re}\left\{\overline{m}+Me^{i(\varphi+2\pi/3)}\right\} 
  \nonumber \\
  m_C & = & {\rm Re}\left\{\overline{m}+Me^{i(\varphi+4\pi/3)}\right\},
  \label{mM}
\end{eqnarray}
where $\overline{m}$ and $Me^{i\varphi}$
are the new order parameters.
The mapping can be inverted by
\begin{eqnarray}
  \overline{m} & = & \frac{\displaystyle m_A+m_B+m_C}{\displaystyle 3}
  \nonumber \\
  Me^{i \varphi} & = & m_A - \overline{m} 
  + i \frac{\displaystyle m_C-m_B}{\displaystyle \sqrt{3}},
\end{eqnarray}
where it is evident that $\overline{m}$ is the average magnetization
and $Me^{i\varphi}$ is an additional (complex) order parameter
characterizing the breaking of translational symmetry.
Several guess solutions have been tried:
a uniform disordered one,
with $\overline{m}$ and $M$ equal to zero ($\varphi$ undefined),
and different symmetry-broken ones, 
with both zero and finite average magnetization $\overline{m}$.
The guess probability distributions have then been obtained
in the hypothesis of statistically independent spins.
First of all the zero field, zero temperature case has been considered.
The disordered solution correctly displays
magnetizations $m_A=m_B=m_C=0$,
pair correlations $c_{AB}=c_{BC}=c_{CA}=-1/3$
and hence internal energy $u=-1$
for any value of $L$
(which is consistent with the fact that only the configurations having
2 spin ``up'' and 1 spin ``down'' on each triangle or vice versa are allowed).
The entropy increases upon increasing $L$
(as displayed in Tab.~\ref{tab:dati})
and converges to the exact zero temperature value~\cite{wannier,houtappel}
$s_{\rm ex} \approx 0.323066$.
A power law behavior of the coherent anomaly type 
(see Ref.~\onlinecite{Suzuki} and references therein)
is observed:
\begin{equation}
  s_L = s_\infty - aL^{-\psi},
\end{equation}
where $s_L$ is the entropy computed with a cluster size equal to $L$.
Discarding the data points with $L=1 \div 3$ 
a least square fit to the above law has been obtained with
$s_\infty \approx 0.323126$ ($\approx s_{\rm ex}$),
$a \approx 0.04647$ and $\psi \approx 1.7512$ (remarkably close to $7/4$),
as shown in Fig.~\ref{fig:entropia}.

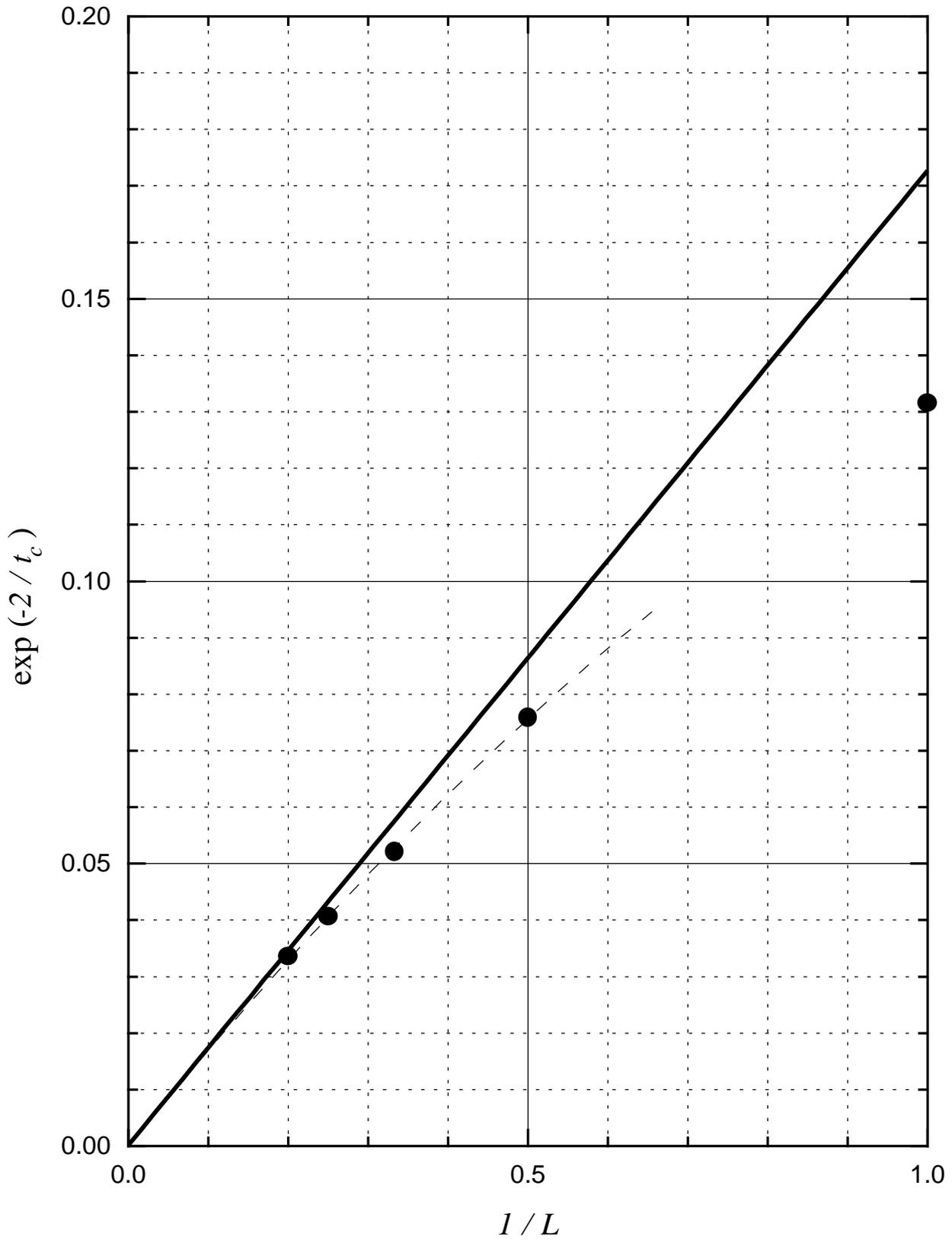
\begin{figure}
\caption{
  The scaled critical temperature $\exp(-2/t_c)$
  is plotted versus the reciprocal cluster size $1/L$.
  Dots represent data obtained by the CVM
  while the solid straight line is 
  a possible asymptotic regime converging to $t_c=0$,
  obtained as the tangent (in $1/L=0$)
  of a parabolic fit (thin dashed line).
  }
\label{fig:tcfit}
\end{figure}

As far as the ordered phases are concerned all guess solutions
turn out to converge to three types only,
which can be distinguished by the value of $\varphi$.
To be more precise
$\varphi$ is exactly $\pi/3$ and $\pi/6$
(for any $L$) for two types of solutions
and seems to approach $\pi/4$
(upon increasing $L$) for a third type.
Taking into account the degeneracy 
induced by the symmetry of the hamiltonian
we actually have
$\varphi=\pi/3+k\pi/3$ (with $k=0,\dots,5$) or 
$\varphi=\pi/6+k\pi/3$ (with $k=0,\dots,5$) 
for the former two types,
and $\varphi \approx \pi/4+k\pi/6$ (with $k=0,\dots,11$)
for the third type.
Particular care is needed in using the NI method
for within the commonly used tolerances
the three types of ordered solutions seem to be all present
for any value of $L$ but,
if precision is increased,
only one or two of the symmetry broken phases
turn out to be real minima of the free energy.
Numerical results are summarized in Tab.~\ref{tab:dati},
where the entropy and the order parameters of each phase
(also the metastable ones)
are displayed as a function of the cluster size $L$.
The alternation between the different ordered solutions is evident.
Also in this case the internal energy does not depend on $L$
and turns out to be $u=-1$,
while a convergence of the entropy towards the exact value
is recovered but no power law can be observed.
On the contrary it is possible to see that the average magnetization 
$\overline{m}$ of each ordered phase
is always very small
(actually vanishing for the phase with $\varphi=\pi/6$)
and approaches zero upon increasing $L$
(see Tab.~\ref{tab:dati}).
The value of $M$ turns out to decrease with $L$, too.
The corresponding states in terms of sublattice magnetizations
can be easily obtained by Eqs.~\ref{mM}.
It is also possible to observe (see Tab.~\ref{tab:dati}) that
the difference of the entropies of the different minima
gets lower and lower upon increasing the cluster size
and this mimics the fact that the model has a continuous rotational symmetry 
(i.e. its free energy is independent of $\varphi$),
which is related to the existence of a field induced 
Kosterlitz-Thouless transition
to a long range ordered phase in the ground state~\cite{Blote1,Blote2}.
Notice that the entropy of the paramagnetic solution
is always lower than that of the ordered solutions
i.e. a (spurious) phase transition is predicted
(at least for any $L$ investigated),
but this is a common feature of CVM approximations
(for instance the hexagon approximation~\cite{kkk} 
predicts a stable phase of the type $\varphi=\pi/6$).
Nevertheless we have verified that the approximate free energy
tends to the exact one (which is known in zero field),
upon increasing $L$.
We have also calculated the critical temperature $t_c$,
limiting the analysis to $L=1 \div 5$,
because computation is very time consuming near critical points.
The transition temperature decreases upon increasing $L$, as expected,
but the convergence towards the exact one (zero in this case)
is not so fast as in other investigations 
based on the CVM~\cite{Kikuchi-Brush}.
Anyway our results seem to be compatible 
with an asymptotic behavior of the following form 
(see Fig.~\ref{fig:tcfit}):
\begin{equation}
  e^{-2/t_c} \propto 1/L
\end{equation}
where $e^{-2/t}$, which in the low temperature limit is proportional
to the correlation length,  is the natural scaling variable
here~\cite{Jacobsen}. 
A more detailed investigation about 
the convergence of the CVM ${\rm B}_{2L}$ approximation hierarchy
towards the continuous symmetry present in the model
and the convergence of the (spurious) transition temperature
towards the exact one
is beyond the scope of the present paper
and is left for future work.

Before concluding this section, devoted to the CVM,
we shall briefly introduce a simple CVM-like approximation,
known as ``cactus'' approximation~\cite{Nagahara-Fujiki-Katsura},
which turns out to be qualitatively correct
for the Ising triangular antiferromagnet.
In the cactus triangle approximation 
the entropy expansion takes into account
only upward- (or downward-) pointing triangles as basic clusters
and hence it can be written as
\begin{equation}
  s = -\sum_{\mbox{\boldmath $\sigma$}_{1,2,3}}
  \rho_{123} \left(\mbox{\boldmath $\sigma$}_{1,2,3}\right)
  \left[
  \ln \rho_{123} \left(\mbox{\boldmath $\sigma$}_{1,2,3}\right)
  - \frac{2}{3} \sum_{i=1}^3 \ln \rho_i \left(\sigma_i\right)
  \right],
\end{equation}
where $\rho_{123} \left(\mbox{\boldmath $\sigma$}_{1,2,3}\right)$
is the triangle probability distribution,
while $\rho_i \left(\sigma_i\right)$
is the site probability distribution,
corresponding to the $i$-th sublattice.
Notice that,
due to the choice of one kind of triangles only
(upward- or downward-pointing),
there is no pair contribution to the CVM entropy.
It is not difficult to prove that
this kind of approximation only predicts a {\em disordered phase}
down to zero temperature (in zero field)
and its entropy turns out to be {\em positive}
($s = \ln (3/2) \approx 0.4055$).
Both characteristics,
even though obtained by a very simple
(and actually not well justified) approximation,
turn out to be qualitatively correct.

\section{The Hard-Spin Mean-Field Theory}

In this section, 
after briefly reviewing the HSMF approximation,
we perform on this basis 
a nearly analytical evaluation of the zero temperature entropy.
This is a new result (exact within the HSMF theory) and, 
though limited to zero temperature,
is an attempt to go beyond approximations
like that proposed by Kabak\c{c}io\={g}lu and coworkers~\cite{berker4}
for the evaluation of HSMF free energy.
Incidentally it is also possible to obtain 
several informations about the model's behavior near zero temperature
in this approximation.

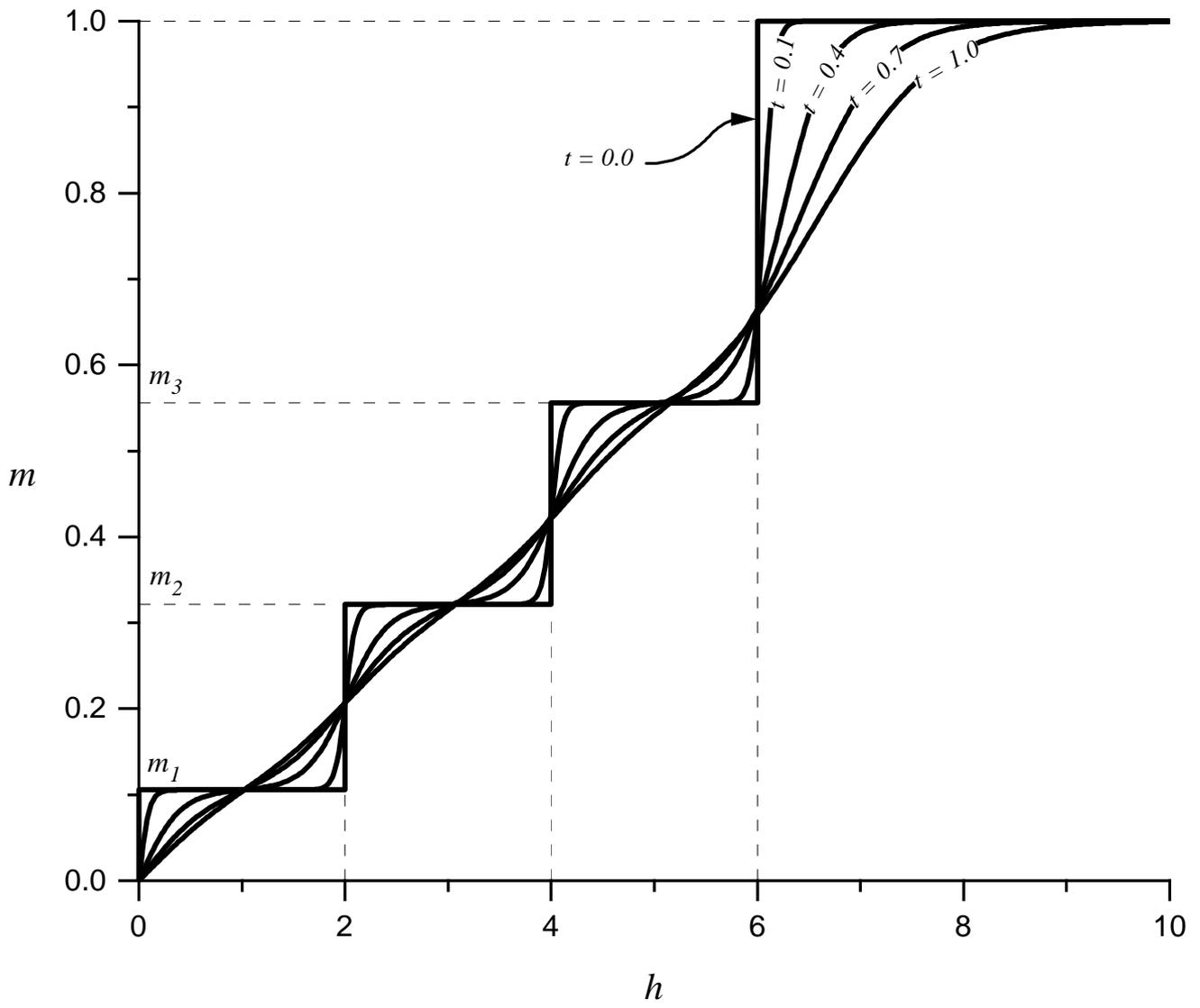
\begin{figure}
\caption{
  Magnetization $m$ vs.\ field $h$
  for the paramagnetic phase
  at different temperatures $t=0.1,0.4,0.7,1.0$
  according to the HSMF theory.
  The limit behavior $t=0$ is also displayed.
  Numerical values of the plateau magnetizations $m_1$, $m_2$, $m_3$ 
  are given by Eqs.~\protect\ref{mplatval}.
  }
\label{fig:magnet1}
\end{figure}

It is easy to show that the following exact relation holds:
\begin{equation}
  \langle \sigma_0 \rangle = 
  \left\langle \tanh \frac{ h - \sum_{r=1}^{6} \sigma_r }{ t } \right\rangle,
  \label{callen_ident}
\end{equation}
where $\sigma_0$ is a spin variable at some site, 
and $\{ \sigma_r \}_{r=1}^{6}$ are its six nearest neighbors.
As usual $\langle \cdot \rangle$ denote thermal average.
The above equation is one of the Callen identities~\cite{Callen}
specialized for the antiferromagnetic model on a triangular lattice.
The classical mean field approximation 
can be derived from Eq.~\ref{callen_ident}
by replacing $\langle \tanh ( \cdot ) \rangle$
by $\tanh \langle \cdot \rangle$.
On the contrary the HSMF approximation consists
in evaluating the right hand side average in Eq.~\ref{callen_ident}
by assuming the nearest neighbor spins $\{ \sigma_r \}_{r=1}^{6}$ 
are statistically independent~\cite{maritan}.
With this assumption 
it is easy to derive from Eq.~\ref{callen_ident}
an equation for the magnetization of a homogeneous phase,
which we are now interested in.
The magnetization $m$ is equal to the thermal average of any spin.
We can then write
\begin{equation}
  m = 
  \langle \sigma_r \rangle = 
  {\rm P} \{ \sigma_r = +1 \} - {\rm P} \{ \sigma_r = -1 \}
  \hspace{1cm} r = 0,1,\dots,6
\end{equation}
and hence
\begin{equation}
  {\rm P} \{ \sigma_r = \pm 1 \} =
  \frac{1 \pm m}{2}
  \hspace{1cm} r = 0,1,\dots,6
  \label{prob_mag}
\end{equation}
where ${\rm P} \{ \cdot \}$ denotes the probability of the event
described within curly braces.
Let $\nu^+$ and $\nu^-$ be the number of nearest neighbor spins
``up'' and ``down'' respectively.
In the hypothesis of statistical independence introduced above
and making use of Eq.~\ref{prob_mag}
it is possible to write the probability distribution of $\nu^+$ and $\nu^-$
by the following formula:
\begin{equation}
  {\rm P} \{ \nu^+ = n \} = {\rm P} \{ \nu^- = 6-n \} = 
  p_n^{(6)} \left( \frac{1+m}{2} \right)
  \hspace{1cm} n=0,\dots,6
\end{equation}
where we have defined the binomial probability distribution
\begin{equation}  
  p_n^{(N)}(x)  \stackrel{\triangle}{=}
  { N \choose n } x^{n} (1-x)^{N-n}.
\end{equation}
Noticing that if $\nu^+ = n$ (and $\nu^- = 6-n$) then
\begin{equation}
  \sum_{r=1}^{6} \sigma_r  =  \nu^+ - \nu^-  =  2(n-3),
\end{equation}
it is finally straightforward to rewrite Eq.~\ref{callen_ident} into
\begin{equation}
  m = \sum_{n=0}^{6} 
  \tanh \frac{ h - 2(n-3) }{ t } \;
  p_n^{(6)} \left( \frac{1+m}{2} \right),
  \label{eqm1}
\end{equation}
which is the equation for $m$ we were looking for.
This equation can be numerically solved
and the magnetization of the homogeneous phase can be computed 
at the given values of temperature $t$ and field $h$.
The equation is evidently invariant under the transformation
$(h \leftrightarrow -h \;,\; m \leftrightarrow -m)$ 
and we have then performed the calculation only for $h \geq 0$
and for several temperatures: 
the results are displayed in Fig.~\ref{fig:magnet1}.
Notice that we have only computed the magnetization
of the homogeneous phase but we have not proved
that it is the stable phase. 
So after calculating the entropy
at $t=0$ and $h=0$ we will have to make sure that the model
does not predict another (inhomogeneous) phase.
The limit magnetization curve at $t=0$, 
displayed in Fig.~\ref{fig:magnet1}, 
is a discontinuous (step) function.
The plateau values of magnetization 
$m_k \; (k=1,2,3)$
can be rigorously computed observing that
\begin{equation}
  \lim_{t \to 0} \; \tanh \frac{ h - 2(n-3) }{ t } = 
  {\rm sgn} \, [ h - 2(n-3) ],
  \label{limtanh}
\end{equation}
where sgn is the sign function, which returns $+1$, $-1$ or $0$
when its argument is positive, negative or zero respectively.
By substituting the above equation into Eq.~\ref{eqm1},
for $2(k-1) < h < 2k \; (k=1,2,3)$ one obtains
\begin{equation}
  m = \sum_{n=0}^{k+2} p_n^{(6)} \left( \frac{1+m}{2} \right)
    - \sum_{n=k+3}^{6} p_n^{(6)} \left( \frac{1+m}{2} \right).
\end{equation}
These are polynomial equations 
which can be written in the more compact form
\begin{equation}
  \frac{1+m}{2} - P_{k+2}^{(6)} \left( \frac{1+m}{2} \right) = 0
  \label{mplateau}
\end{equation}
by defining the cumulative distribution
\begin{equation}
  P_n^{(N)}(x)  \stackrel{\triangle}{=}
  \sum_{n'=0}^{n} p_{n'}^{(N)}(x)
  \label{poly}
\end{equation}
and making use of Newton's binomial formula.
It is now easy to find all the solutions 
of the polynomial (6-th degree) equations~\ref{mplateau}
by common numerical routines; 
$m_k$ is, out of the solutions of the $k$-th equation,
the only one in the real interval $[-1, 1]$:
\begin{eqnarray}
  m_1 & \approx & 0.1056 \\ \nonumber
  m_2 & \approx & 0.3213 \\ \nonumber
  m_3 & \approx & 0.5562.
  \label{mplatval}
\end{eqnarray}
It is also easy to see that for $h>6$ Eq.~\ref{eqm1}
simply reduces to $m=1$.
The information about the magnetization plateau values
is already sufficient for the zero temperature zero field internal energy 
$u\left(t=0,m=0\right)$ to be evaluated.
The following thermodynamic identity must be used:
\begin{equation}
  u(t=0,m=0) = u(t=0,m=1) - \int_0^1 h(t=0,m) dm.
\end{equation}
The integral can be easily evaluated 
for $h\left(t=0,m\right)$, 
implicitly defined via Eq.~\ref{eqm1} in the limit $t \to 0$,
is a step function:
\begin{equation}
  h(t=0,m) = 2k  \hspace{1cm}  \forall m \in (m_k, m_{k+1}),
  \label{h0m}
\end{equation}
where $k=0,\dots,3$ and $m_0=-m_1$ and $m_4=1$.
Moreover $u\left(t=0,m=1\right) = 3$ for 
if $m=1$ then the spin pair correlation is $c=1$ 
and $u=3c$ in the triangular antiferromagnetic model.
We finally obtain
\begin{equation}
  u(t=0,m=0) = 2(m_1+m_2+m_3) - 3 \approx -1.0339.
\end{equation}
It is a remarkable fact that 
the pair correlation is then 
$c(t=0,m=0) \approx -0.3446$, 
which is quite close to the exact value $-1/3$, 
but it does not respect the compatibility condition
$c \geq -1/3$.

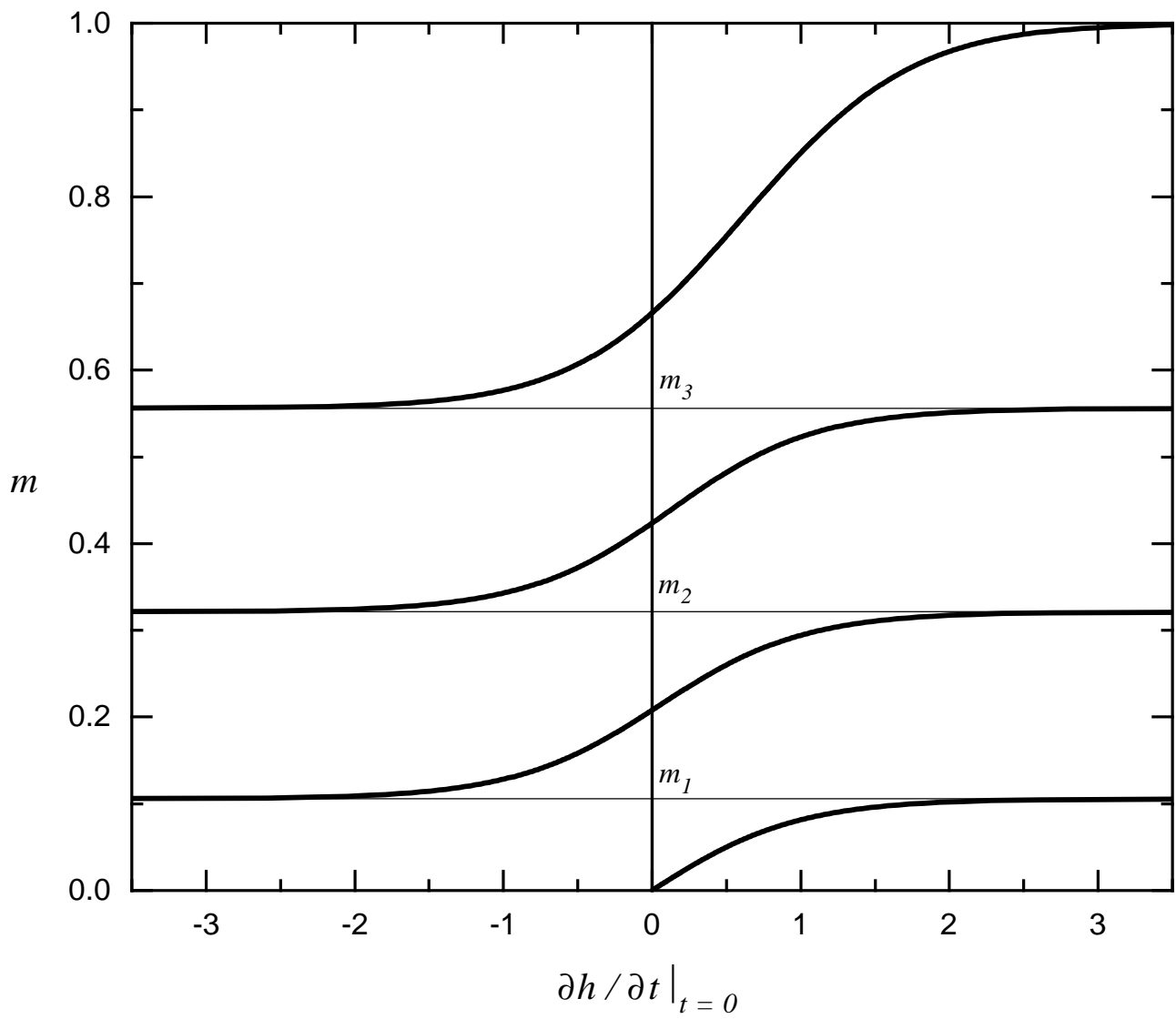
\begin{figure}
\caption{
  Partial derivative of the field $h$ 
  with respect to temperature $t$ in $t=0$ 
  as a function of magnetization $m$.
  In analogy with Fig.~\protect\ref{fig:magnet1},
  $m$ is reported on the vertical axis
  (and $\partial h / \partial t$ on the horizontal axis).
  Numerical values of $m_1$, $m_2$, $m_3$ 
  are given by Eqs.~\protect\ref{mplatval}.
  }
\label{fig:dhdt}
\end{figure}

In order to evaluate 
the zero temperature zero field entropy $s\left(t=0,m=0\right)$
we need further manipulations.
Again we will make use of a thermodynamic identity:
\begin{equation}
  s(t=0,m=0) = s(t=0,m=1) + 
  \int_0^1 \frac{\partial h}{\partial t} (t=0,m) dm,
  \label{entropia}
\end{equation}
where the entropy of the saturated system 
$s\left(t=0,m=1\right)$ vanishes.
As far as the evaluation of the integral is concerned
we have to derive an explicit expression
for the partial derivative of $h(t,m)$
with respect to $t$ in $t=0$.
To do that let us consider Eq.~\ref{eqm1} and, for $k=0,\dots,3$,
solve it with respect to $(h-2k)/t$.
With a bit of algebra we can write
\begin{equation}
  \frac{ h(t,m) - 2k }{ t } = 
  \tanh^{-1} \frac
  {\displaystyle 
    m - \sum_{\stackrel{\scriptstyle n=0}{n \neq k+3}}^6
    \tanh \frac{h(t,m)-2(n-3)}{t} \;
    p_n    ^{(6)} \left( \frac{1+m}{2} \right) }
  {\displaystyle 
    p_{k+3}^{(6)} \left( \frac{1+m}{2} \right) }.
\end{equation}
Let us now take the limit $t \to 0$
of each side of the above equation.
Making use of Eq.~\ref{limtanh} and \ref{h0m}
we easily obtain
\begin{eqnarray}
  \lim_{t \to 0} \frac{ h(t,m) - h(t=0,m) }{ t } & = &
  \tanh^{-1} \frac
  {\displaystyle 
    m -
    \sum_{n=0}^{k+2} p_n^{(6)} \left( \frac{1+m}{2} \right) +
    \sum_{n=k+4}^{6} p_n^{(6)} \left( \frac{1+m}{2} \right) }
  {\displaystyle 
    p_{k+3}^{(6)} \left( \frac{1+m}{2} \right) } \nonumber \\ 
  & & \forall m \in (m_k, m_{k+1}).
\end{eqnarray}
Notice that the left hand side of this equation
is just the definition of the partial derivative
we were looking for
but on the right hand side
a different expression is obtained 
for each interval $(m_k, m_{k+1})$.
Expanding the inverse hyperbolic tangent
and introducing definition~(\ref{poly}),
we finally have
\begin{eqnarray}
  \frac{\partial h}{\partial t} (t=0,m) & = &
  \frac{1}{2} \ln \frac
  {\displaystyle 
    \left[ \frac{1+m}{2} - P_{k+2}^{(6)} \left( \frac{1+m}{2} \right) \right] } 
  {\displaystyle 
   -\left[ \frac{1+m}{2} - P_{k+3}^{(6)} \left( \frac{1+m}{2} \right) \right] }
  \nonumber \\ & & \forall m \in (m_k, m_{k+1}).
  \label{dhdt} 
\end{eqnarray}
The above result is graphically reported in Fig.~\ref{fig:dhdt}.
Due to singularities
it is not easy to perform an integration of such a function
by common numerical quadrature algorithms,
but the form of the function permits the following manipulations.
Substituting Eq.~\ref{dhdt} into Eq.~\ref{entropia}
and making the change of variable $x = (1+m)/2$
it is possible to write
\begin{equation}
  s(t=0,m=0) = \sum_{k=0}^3
  \left[
  \int_{x_k}^{x_{k+1}} \ln \left| x - P_{k+2}^{(6)}(x) \right| dx -
  \int_{x_k}^{x_{k+1}} \ln \left| x - P_{k+3}^{(6)}(x) \right| dx
  \right],
  \label{entropia2}
\end{equation}
where $x_k = (1+m_k)/2$ for $k=1,\dots,4$ and $x_0=1/2$.
It has been possible to introduce absolute values
because only one solution of Eq.~\ref{mplateau}
lies in the physically meaningful region $m \in [-1,1]$
and hence in that interval the following relation holds:
\begin{equation}
  \frac{1+m}{2} - P_{k+2}^{(6)} \left( \frac{1+m}{2} \right) > 0
  \hspace{3mm} \Longleftrightarrow \hspace{3mm}
  m > m_k,
\end{equation}
where $k=0,\dots,4$.
The integrals of Eq.~\ref{entropia2} can now be computed
by means of the following formula:
\begin{equation}
  \int \ln |\wp(x)| dx = 
  {\rm Re} \left\{ x \ln a_g + \sum_{i=1}^g
  (x-z_i) \left[ \ln (x-z_i) - 1 \right] \right\}
  + {\rm const.},
\end{equation}
where $\wp(x)$ is any ($g$-th degree) polynomial, 
$z_i$ is its $i$-th (in general complex) root
and $a_g$ is the highest power coefficient.
The numerical problem is then reduced to the computation
of polynomial roots.
We have obtained
\begin{equation}
  s(t=0,m=0) \approx 0.3869,
\end{equation}
which is positive and not so far from the exact value.

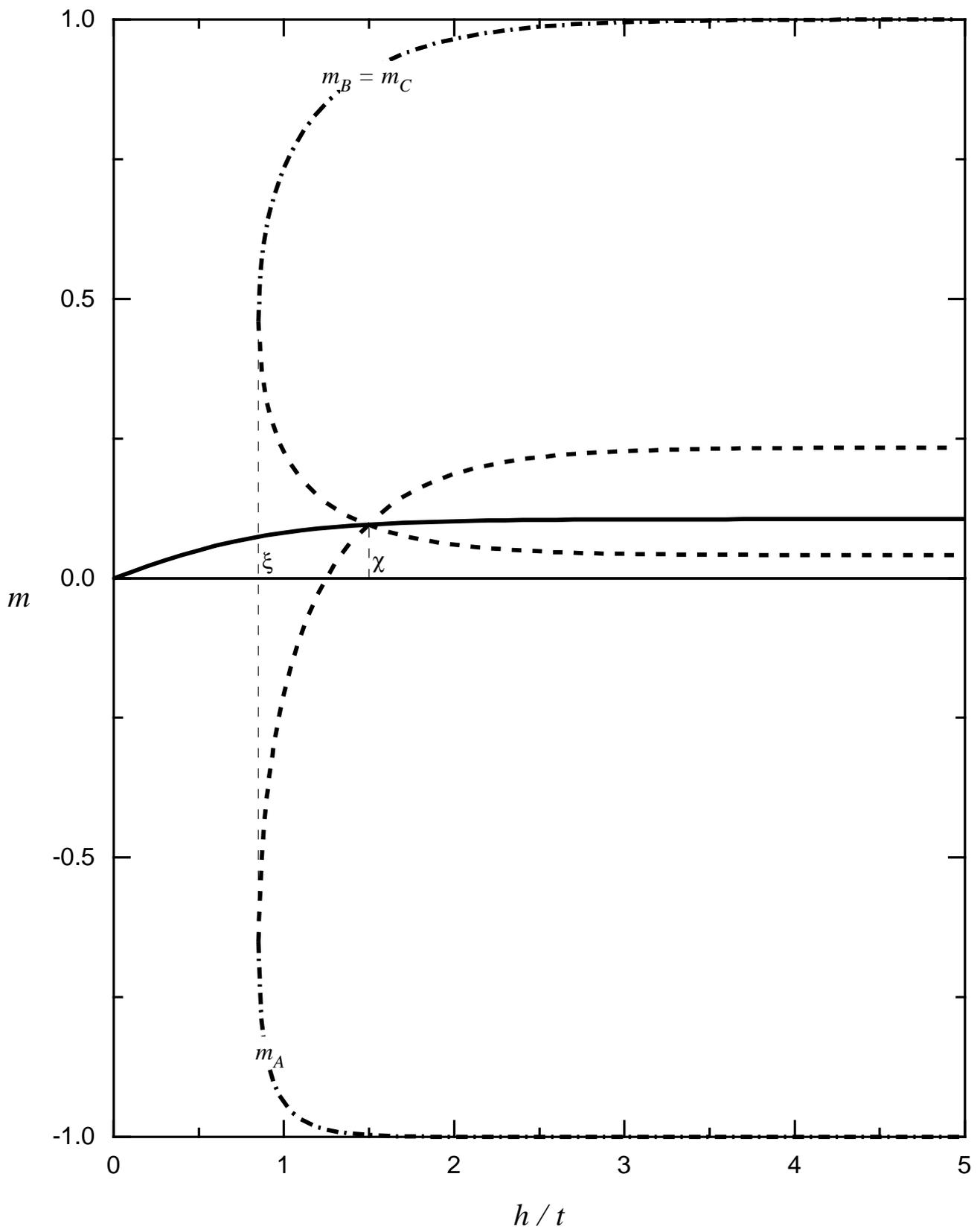
\begin{figure}
\caption{
  Magnetization $m$ as a function of $h/t$
  in the asymptotic regime $t \to 0$
  (in the neighborhood of $h=0$).
  For both symmetry-broken solutions
  the magnetizations of the two non-equivalent sublattices
  are denoted by $m_B=m_C$ and $m_A$;
  the two solutions become critical at $h/t=\xi$.
  The uniform magnetization (solid line)
  crosses that of one broken-symmetry solution (dashed lines) 
  at $h/t=\chi$.
  The other broken-symmetry solution 
  is represented by dash-dotted lines.
  }
\label{fig:magnet2}
\end{figure}

As a final check
we have examined the possibility of symmetry-broken phases,
as usual limiting our investigation to the case of a tripartite lattice.
The unknowns of this problem are the magnetizations
of the three sublattices,
which we have called as usual $m_A$, $m_B$ and $m_C$.
Along the lines of the derivation of the equation for $m$
in the uniform case,
and observing that if a site is of type $A$
its nearest neighbors are three of type $B$ and three of type $C$
and the same for all permutations of $A$, $B$, $C$,
it is straightforward to write the following three equations:
\begin{equation}
  m_a = \sum_{n'=0}^{3} \sum_{n''=0}^{3} 
  \tanh \frac{ h - 2(n'+n''-3) }{ t } \;
  p_{n'}^{(3)} \left( \frac{1+m_b}{2} \right)
  p_{n''}^{(3)} \left( \frac{1+m_c}{2} \right), 
  \label{eqm2}
\end{equation}
with $(a\;b\;c) = (A\;B\;C), (B\;C\;A), (C\;A\;B)$.
A numerical solution of the system has shown that
two symmetry-broken solutions with two equivalent sublattices
(analogous to those obtained by Bethe approximation)
exist in the vicinity of $t=0$.
As displayed in Fig.~\ref{fig:magnet2},
and in analogy to what happens for Bethe approximation,
symmetry-broken solutions are not present at $h=0$
because they become critical
in two simmetrical points $h/t = \pm \xi$.
This result, as well as the existence of two points ($h/t = \pm \chi$)
in which the uniform solution crosses one simmetry-broken solution,
has been extensively pointed out in Ref.~\onlinecite{berker4}.
Notice that in Fig.~\ref{fig:magnet2}
magnetizations are functions of $h/t$ only.
This is due to the form of Eqs.~\ref{eqm2} in which,
for $t \to 0$ and in a region of $h$ close to zero,
only one hyperbolic tangent
(the one whose argument is $h/t$)
is significantly different from $\pm 1$.
In this way we can be sure that
the one displayed in Fig.~\ref{fig:magnet2}
is the asymptotic behavior of magnetizations for $t \to 0$
and hence no symmetry-broken solution can appear at $h=0$
for any arbitrarily small temperature value.

\section{Conclusions}
We have investigated several mean-field like approximations 
for the antiferromagnetic Ising model on the triangular lattice. 
We have shown that 
the paramagnetic phase predicted by the BP approximation 
is unphysical in a low temperature, low field region since, 
due to the neglect of frustration effects, 
the approximation predicts as the ground state 
a state which cannot exist on the triangular lattice, 
and hence has negative entropy. 
In the case of the CVM we have investigated 
some of the convergence properties of the $B_{2L}$ series of approximations, 
showing that the zero field zero temperature entropy 
of the (metastable) disordered phase
converges to the exact value with a power law 
of the coherent anomaly type, $s_L = s_{\rm ex} - a L^{-\psi}$, 
with $\psi \approx 7/4$. 
The CVM predicts a variety of ordered ground states
(and hence a spurious phase transition)
whose symmetry depends on the cluster size
and a possible interpretation of this behavior
has been suggested.
Subsequently the critical temperature has been evaluated,
limiting the analysis to $L=5$,
due to the well-known computational difficulties near critical points.
The transition temperature turns out to decrease with $L$,
as expected, although very slowly.
We have also mentioned that the cactus triangle approximation 
behaves in a qualitatively correct way. 
Finally, we have considered the HSMF theory, 
calculating for the first time 
the zero field zero temperature entropy and internal energy 
by means of a well-defined procedure.



\begin{table}
  \caption{
    CVM entropy expansion coefficients 
    for Kikuchi's ${\rm B}_{2L}$ hierarchy on a triangular lattice:
    $\gamma$ denotes a cluster class
    (${\sf W}$, ${\sf N}$, ${\sf O}$, ${\sf E}$),
    $a_\gamma$ is obtained by Moebius inversion~\protect\cite{An}
    and $N_\gamma / N$ is the number of $\gamma$-clusters per lattice site.
  }
  \label{tab:coefficienti}
  \begin{tabular}{|c||c|c|}
    \hline
    $\gamma$ & $a_\gamma$ & $N_\gamma / N$ \\
    \hline \hline
    ${\sf W}$ & $+1$ & $2$ \\
    ${\sf N}$ & $-1$ & $2$ \\
    ${\sf O}$ & $-1$ & $1$ \\
    ${\sf E}$ & $+1$ & $1$ \\
    \hline    
  \end{tabular}
\end{table}

\begin{table}
  \caption{
  Results obtained by the CVM
  for different cluster sizes (first column).
  The second and third columns contain respectively 
  the argument $\varphi$ 
  (discriminating the type of symmetry breaking,
  $\varphi$ is undefined for the disordered phase)
  and the modulus $M$ of the complex order parameter,
  whereas the fourth column contains 
  the average magnetization $\overline{m}$.
  The fifth column contains the (reduced) entropy $s$:
  for each value of the cluster size $L$ 
  the different solutions are ordered on decreasing entropy
  (the first is the stable one).
  For $L=9$ only the disordered phase has been investigated.
  }
  \label{tab:dati}
  \begin{tabular}{|c||l|rr|l|}
    \hline
    $L$ & $\varphi$ (deg.) & $M$ & $\overline{m}$ & $s$ \\
    \hline \hline
    $1$ & $30.0000$ & $0.89810$ & $0.00000$ & $0.3095711$ \\ 
    $ $ & $60.0000$ & $0.86914$ & $0.00533$ & $0.3051012$ \\ 
    $ $ & $-      $ & $0.00000$ & $0.00000$ & $0.2876821$ \\ 
    \hline
    $2$ & $60.0000$ & $0.79214$ & $0.00252$ & $0.3177109$ \\
    $ $ & $-      $ & $0.00000$ & $0.00000$ & $0.3095711$ \\
    \hline
    $3$ & $30.0000$ & $0.70585$ & $0.00000$ & $0.3207863$ \\
    $ $ & $-      $ & $0.00000$ & $0.00000$ & $0.3164601$ \\
    \hline
    $4$ & $47.2021$ & $0.68361$ & $0.00041$ & $0.3210938$ \\
    $ $ & $60.0000$ & $0.67748$ & $0.00045$ & $0.3210694$ \\
    $ $ & $-      $ & $0.00000$ & $0.00000$ & $0.3190285$ \\
    \hline
    $5$ & $60.0000$ & $0.65295$ & $0.00046$ & $0.3217165$ \\    
    $ $ & $30.0000$ & $0.65240$ & $0.00000$ & $0.3217141$ \\
    $ $ & $-      $ & $0.00000$ & $0.00000$ & $0.3203469$ \\    
    \hline
    $6$ & $45.1102$ & $0.62727$ & $0.00023$ & $0.3221726$ \\    
    $ $ & $-      $ & $0.00000$ & $0.00000$ & $0.3211130$ \\    
    \hline
    $7$ & $30.0000$ & $0.59927$ & $0.00000$ & $0.3223160$ \\    
    $ $ & $60.0000$ & $0.59818$ & $0.00018$ & $0.3223155$ \\    
    $ $ & $-      $ & $0.00000$ & $0.00000$ & $0.3215873$ \\    
    \hline
    $8$ & $60.0000$ & $0.58114$ & $0.00016$ & $0.3224757$ \\
    $8$ & $30.0000$ & $0.58017$ & $0.00000$ & $0.3224749$ \\    
    $ $ & $-      $ & $0.00000$ & $0.00000$ & $0.3219084$ \\    
    \hline
    $9$ & $-      $ & $0.00000$ & $0.00000$ & $0.3221350 $ \\    
    \hline    
  \end{tabular}
\end{table}



\begin{figure}
\caption{
  In the field-temperature $(h,t)$ plane
  the loci of points in which the BP approximation predicts 
  $s=0$ (thick dashed line) and $c=-1/3$ (thick solid line)
  are displayed.
  Some contour lines of the NN pair correlation $c$
  are also reported (thin lines).
  }
\label{fig:ccost}
\end{figure}


\begin{figure}
\caption{
  Basic clusters of the ${\rm B}_{2L}$ hierarchy on a triangular lattice.
  Numerical ordering of sites is that used in formulas.
  }
\label{fig:indici}
\end{figure}

\begin{figure}
\caption{
  Cluster involved in the ${\rm B}_{2L}$ entropy expansion
  for a triangular lattice
  split into 3 non equivalent sublattices ($A$, $B$, $C$).
  Due to lattice splitting
  each cluster class (${\sf W}$, ${\sf N}$, ${\sf O}$, ${\sf E}$)
  is split into 3 subclasses or ``types'' ($^{(1)}$, $^{(2)}$, $^{(3)}$)
  with the same shapes but different probability distributions.
  }
\label{fig:cluster}
\end{figure}

\end{document}